\documentclass[12pt,letterpaper]{article}

\parskip=\medskipamount            
\def\7#1#2{\mathop{\null#2}\limits^{#1}}        

\def\beee{\begin{equation}}
\def\eeee{\end{equation}}
\def\dggg{^{\dagger}}
\begin{document}

\bibliographystyle{unsrt}
\begin{center}
\textbf{FROM WIGNER'S SUPERMULTIPLET THEORY\\ TO QUANTUM CHROMODYNAMICS}\\
\end{center}
O.W. Greenberg\footnote{email address, owgreen@physics.umd.edu.}\\
Center for Theoretical Physics\\
Department of Physics \\
University of Maryland\\
College Park, MD~~20742-4111\\
University of Maryland Preprint PP-03-028\\
\begin{center}
Talk given at the Wigner Centennial Conference, Pecs, Hungary, July 2002.\\
To appear in the proceedings of the conference.\\
\end{center}
\begin{abstract}

The breadth of Eugene Wigner's interests and contributions is amazing
and humbling. At different times in his life he did seminal work
in areas as diverse as pure mathematics and chemical engineering. His
seminal research in physics is, of course, the best known. In this talk
I first describe Wigner's supermultiplet theory of 1936 using the 
approximate symmetry of the nuclear Hamiltonian under a combined 
spin-isospin symmetry to describe the spectroscopy of stable nuclei 
up to about the nucleus molybdenum. I then show how Wigner's ideas 
of 1936 have had far reaching and unexpected implications: his ideas 
led to the discovery of the color degree of freedom for quarks and 
to the symmetric quark model of baryons which is the basis of baryon 
spectroscopy. I conclude by pointing out that the color degree of 
freedom, made into a local symmetry using Yang-Mills theory, leads 
to the gauge theory of color, quantum chromodynamics, which is our 
present theory of the strong interactions.
\end{abstract}

I am very happy to participate in this Centennial Conference to remember
and honor the life and work of Eugene Wigner. I start by giving some
reminiscences of my contacts with Wigner. When I was a first year graduate
student I approached Wigner at the daily afternoon tea in Fine Hall to ask
a question. I don't remember the question, but I do remember the kindness
with which Wigner treated me. He asked me to join him in his office, which
was also in Fine Hall, and gave me a careful and full answer to my 
question. Once on a winter day I saw Wigner, who was slight in physique,
almost blown in by the wind as he entered the side door of the passage that
connected Palmer Lab with Fine Hall. I had the pleasure to take two courses
with Wigner during my studies, Kinetic Theory and Group Theory. 

Wigner was unfailingly polite. When Wigner had to leave a seminar at
at the Institute for Advanced Study before it had ended, he would carefully
and quietly gather his belongings and try to tiptoe out of the room.
Generally this took some time during which all eyes were on him. His
attempt to be unobtrusive had the opposite effect. On one occasion, Wigner
asked several questions of one of his nuclear physics graduate students
during his seminar. His student was impatient and answered grudgingly.
Wigner shrugged, stopped asking questions and said ``I don't want to slow
down the progress of science.'' 

Among the historic events with which Wigner was involved, two 
have not been mentioned here, so I will recall them now. One is that
Wigner joined Leo
Szilard in driving to Long Island, where Einstein was sailing, to ask him
to write to President Franklin Roosevelt to make him aware of the potential
of a nuclear weapon. The other is that Wigner had the foresight to bring
a bottle of chianti to the squash court under Stagg Field at the University
of Chicago to celebrate the nuclear reactor built under the direction of
Enrico Fermi becoming critical for the first time. All present signed the
wicker basket holding the bottle. I don't know where this historic basket
is now.

In the 
summer of 1962 there was a NATO summer school organized by Feza G\"ursey
at what then was called the Robert College in Bebek, Turkey outside of
Istanbul. The proceedings are still worth reading.\cite{nat} 
Wigner gave a series of lectures on the 
representations of the inhomogenous Lorentz group of relevance to 
elementary particles, with particular emphasis on the extension of the 
group to take account of the discrete elements, parity, time reversal, and
spacetime reversal.\cite{wig} Of course Wigner himself had introduced these discrete
synnetries into quantum mechanics. In these lectures, Wigner found
some unusual representations in which the dimension of the representations
of the discrete elements is doubled. As the senior American physicist, 
Wigner felt responsible to provide a report on the summer school for 
publication in Physics Today. He asked me to write a first draft of the
report, which duly appeared.\cite{phy} After leaving Bebek we both traveled to Trieste, where 
lectures were being given at what became the International Center for
Theoretical Physics founded by Abdus Salam. Among the lecturers was
Julian Schwinger, who gave his solution of two-dimensional quantum
electrodynamics. Both Wigner and I had to leave early one morning to fly
to Venice. We were out on the road by the hotel, cold and hungry, waiting
for a late cab to arrive. At the airport there was no time to eat. We were
flying in an old DC3, which lands on its tail and wing wheels. I found some
old sticks of chewing gum and shared them with Wigner, so we could get
some calories into our bodies.  

Now I turn to the physics part of my talk. Wigner was the first to combine
spacetime and internal symmetries into a larger symmetry group whose 
approximate validity leads to new predictions about measurable quantities.
In Wigner's paper of 1936\cite{wig2} he discusses the various 
types of nuclear forces,
space dependent forces, space and spin dependent forces, space and isospin
dependent forces, and finally, space, spin and isospin dependent forces.
Bear in mind that Werner Heisenberg had introduced the concept of isospin,
that the proton and neutron can be considered as members of an SU(2)
doublet, the nucleon, just four years before, 
in 1932. The spin states of a spin 1/2 particle can
also be considered as an SU(2) doublet. Wigner's seminal idea was to combine
the two groups into their associative covering group, SU(4), and to
explore the consequences of an approximate SU(4) spin-isospin symmetry.
Thus Wigner gave the first example of a symmetry that combined spacetime
and internal symmetries. 
What Wigner did was to combine the spin-1/2 doublet
\beee
\left(\frac{1}{2}\right) \sim 
\left( \begin{array}{c} \uparrow \\  \downarrow  \end{array} \right)
\sim 2_S~~ \rm{in} ~~SU(2)_S
\eeee
with the isospin-1/2 doublet
\beee
\left(N\right) \sim 
\left( \begin{array}{c} p \\ n \\  \end{array} \right)
\sim 2_I~~ \rm{in} ~~SU(2)_I
\eeee
to get a quartet in the larger group
\beee
\left( \begin{array}{c} p \uparrow \\ p \downarrow \\  n \uparrow
\\ n \downarrow \end{array} \right) \sim 4_{SI} ~~ \rm{in} ~~SU(4)_{SI}.
\eeee
One can reverse the process and decompose the quartet into a product of
doublets
\beee
4_{SI} \rightarrow 2_S \otimes 2_I ~~\rm{under} ~~SU(4)_{SI} 
\rightarrow SU(2)_S \otimes SU(2)_I.
\eeee
Wigner showed that his larger symmetry gave a good fit to nuclear states
up to molybdenum. 

Here we have Wigner as the master of the use of group theory in physics.
On the one hand, he used powerful theorems due to G. Frobenius (whom I
have heard him call ``old man Frobenius''), E. Cartan, I. Schur and H.
Weyl. On the other hand, Wigner as a profound physicist used the group
theory results to confront expermental data.

The next step in the story of the fruits of Wigner's idea to combine spin
and isospin symmetry was taken by
Feza G\"ursey and Luigi A. Radicati\cite{gur} in 1964, soon after the 
quark model was
introduced independently by Murray Gell-Mann\cite{gel} and by
George Zweig\cite{zwe}. 
G\"ursey and
Radicati transplanted Wigner's idea to elementary particle physics using
the $SU(2)_S$ symmetry and what now we call the flavor symmetry $SU(3)_F$
and combining them to the associative covering group $SU(6)_{SF}$. Their
construction was closely analogous to Wigner's work of 1936. We have
\beee
\left(\frac{1}{2}\right) \sim 
\left( \begin{array}{c} \uparrow \\  \downarrow  \end{array} \right)
\sim 2_S~~ \rm{in} ~~SU(2)_S
\eeee
combined with the flavor-triplet
\beee
\left(q\right) \sim 
\left( \begin{array}{c} u \\ d \\ s \end{array} \right)
\sim 3_F~~ \rm{in} ~~SU(3)_F
\eeee
to get a sextet in the larger group
\beee
\left( q \frac{1}{2} \right) \sim \left( \begin{array}{c} u \uparrow \\ u 
\downarrow \\ d \uparrow
\\ d \downarrow \\ s \uparrow \\ s \downarrow \end{array} \right) 
\sim 6_{SF} ~~ \rm{in} ~~SU(6)_{SF}.
\eeee
Again the larger representation decomposes to a product of representations
of each of the constituent groups,
\beee
6_{SF} \rightarrow 2_S \otimes 3_F ~~\rm{under}~~ SU(6)_{SF} 
\rightarrow SU(2)_S \otimes SU(3)_F.
\eeee

Great skepticism about quarks reigned in the physics community. 
Fractionally particles had never been seen; indeed they have not been
seen to this day. Gell-Mann himself was ambigous about the reality of 
quarks.

The $SU(6)_{SF}$ theory had some stricking successes: the lowest-mass
mesons, the pseudoscalar
octet and the vector nonet, fit precisely into a $35$ + $1$ of 
$SU(6)_{SF}$. The lowest-mass baryons fit exactly into a $56$ of
$SU(6)_{SF}$. G\"ursey and Radicati derived mass formulas for each 
supermultiplet and these mass formulas agreed well with data. For the 
mesons,
\beee
q \bar{q} \sim 6 \otimes 6^{\star} = 1 ~+
~35~~\rm{in} ~~SU(6)_{SF}
\eeee
\beee
1 \rightarrow (1,0);~~35 \rightarrow (8,0) ~+~(1+8,1) ~~\rm{under}~~
SU(3)_{FS} \otimes SU(2)_S.
\eeee
There is
nothing to raise concern about the placement of the mesons in the $SU(6)$
theory.

The baryons are a different story. Since three quarks are bound in a
baryon (in what we now call the ``constituent'' quark model), the 
possibilities are
\beee
qqq \sim 6 \otimes 6 \otimes 6 = 56 ~+ 70 ~+70~+ 20.
\eeee
The $56$ is totally symmetric under permutations of the quarks;
the $70$'s have mixed symmetry and the $20$ is totally antisymmetric.

The spin-statistics theorem\cite{pau} 
states that integer-spin particles form
symmetric states under permutations and odd-half-integer-spin particles
form antisymmetric states. Since the quarks are assigned spin-1/2 in
order to give the odd-half-integer spin of the baryons, the spin-statistics
theorem requires the quarks to be in the $20$. Bunji Sakita\cite{sak} made 
this
choice. As we will see, both the observed spectrum of lowlying baryons and
the ratio of the magnetic moments of the proton and neutron rule out
this assignment. G\"ursey and Radicati chose the $56$, which fits the
baryons beautifully. This choice, symmetric under permutations, violates
the spin-statistics theorem and presents a paradox: quarks as spin-1/2
particles should be fermions and should obey the Pauli exclusion principle,
yet the quarks are symmetric under permutations in the $56$. This paradox
compounded the problems of the quark model and greatly increased the
reluctance of the physics community to accept the quark model. Unobserved
fractionally charged particles were bad enough. Particles that violated
the exclusion principle were just too incredible.

Nontheless, in addition to the success of the $SU(6)$ theory for the 
lowlying baryons, there was another striking result that agreed with
experiment. M.A.B. Baqi B\'eg, Benjamin W. Lee and Abraham Pais\cite{beg} 
and, independently, Sakita\cite{sak2} calculated the ratio of the magnetic
moments of the proton and neutron and found the simple result
\beee
\frac{\mu_p}{\mu_n}=-\frac{3}{2}
\eeee
using pure $SU(6)$ group theory. This result agrees with experiment to
within $3\%$! All previous calculations using meson cloud effects had
failed utterly. Nobody had realized that the ratio had such a simple
value. The magnetic moment calculation gave heart to those who believed
that the $SU(6)$ theory was on the right track.

The first solution to the spin-statistics paradox was provided by 
parastatistics.
 Parastatistics had been introduced by H.S. Green\cite{green} in
1953. Green noticed that the commutation relations between the number
operator and the annihilation and creation operators are the same for both
bosons and fermions,
\beee
[n_k, a\dggg_l]_-=\delta_{kl}a\dggg_l, ~~[n_k, a_l]_-=-\delta_{kl}a_l
\eeee
The number operator can be written
\beee
n_k=(1/2)[a\dggg_k, a_k]_{\pm}+ {\rm const},
\eeee
where the anticommutator (commutator) is for the Bose (Fermi) case. 
He realized that he could generalize each of these
types of particle statistics to a family, labeled by the order $p$, where
$p=1,2, 3,\cdots$ and, in each case, $p=1$ is the usual statistics. 
If these
expressions are inserted in the number operator-creation 
operator commutation
relation, the resulting relation is \emph{trilinear} 
in the annihilation and creation operators.  Polarizing the number 
operator to
get the transition operator $n_{kl}$ which annihilates a free particle 
in state
$k$ and creates one in state $l$ leads to Green's trilinear commutation relation
for his parabose and parafermi statistics,
\beee
[[a\dggg_k, a_l]_{\pm}, a\dggg_m]_-=2\delta_{lm}a\dggg_k
\eeee
Since these rules are trilinear, the usual vacuum condition,
\beee
a_k|0\rangle=0,
\eeee
does not suffice to allow calculation of matrix elements of the $a$'s and
$a\dggg$'s; a condition on one-particle states must be added,
\beee
a_k a\dggg_l|0\rangle=\delta_{kl}|0\rangle.
\eeee

Green found an infinite set of solutions of his commutation rules, 
one for each 
integer, by giving an ansatz which he expressed in terms of Bose and Fermi
operators.  Let
\beee
a_k\dggg=\sum_{p=1}^n b_k^{(\alpha) \dagger},~~a_k=\sum_{p=1}^n b_k^{(\alpha)},
\eeee
and let the $b_k^{(\alpha)}$ and $b_k^{(\beta) \dagger}$ 
be Bose (Fermi) operators
for $\alpha=\beta$ but anticommute (commute) for $\alpha \neq \beta$ for the 
``parabose'' (``parafermi'') cases.  This ansatz clearly satisfies Green's
relation.  The integer $p$ is the order of the parastatistics.  The physical
interpretation of $p$ is that, for parabosons, $p$ is the maximum number of
particles that can occupy an antisymmetric state; for parafermions, $p$
is the maximum number of particles that can occupy a symmetric state (in
particular, the maximum number which can occupy the same state).  The case $p=1$
corresponds to the usual Bose or Fermi statistics.
Later, Messiah
and I\cite{gremess} proved that Green's ansatz gives all Fock-like 
solutions of
Green's commutation rules.  Local observables have a form analogous to 
the usual
ones; for example, the local current for a spin-1/2 theory is 
$j_{\mu}=(1/2)[\bar{\psi}(x), \psi(x)]_-$.  From Green's ansatz, it is 
clear
that the squares of all norms of states are positive, since sums of Bose or
Fermi operators give positive norms.  Thus parastatistics gives a set of
orthodox theories.  Parastatistics is one of the
possibilities found by Doplicher, Haag and Roberts\cite{dop} in a 
general study of
particle statistics using algebraic field theory methods.  
A good review of this
work is in Haag's book\cite{haa}.

Note that Wigner\cite{wig3} anticipated parastatistics in his study of 
generalizations of the harmonic oscillator. His case corresponds to 
parabose statistics of order 2 for a single oscillator.

The spin-statistics theorem generalizes in the context of parastatistics.
The theorem becomes \emph{Given the choice between parabose and parafermi
statistics}, parabose particles must have integer spin and parafermi 
particles must have odd-half-integer spin\cite{del}.

My colleagues at Maryland had asked me to invite G\"ursey, who in
the summer of 1964, was in
Brookhaven, to give a seminar at Maryland. When I
called him, he said that he and Radicati had just finished some interesting
work that would soon appear in Physical Review Letters. I was to be on
leave at the Institute for Advanced Study. When I arrived there was great
interest in the $SU(6)$ theory. Ben Lee showed me the magnetic moment 
calculation which really convinced me that the $SU(6)$ theory was correct.
Because I had been working on parastatistics with 
A.M.L. Messiah at Saclay, I immediately realized that if the quarks obeyed
parafermi statistics of order 3 the contradiction with the spin-statistics
theory would be removed.  

I was able to show that three parafermi quarks of order three can be in
a state that is totally symmetric under permutations of the visible degrees
of freedom, space, spin and unitary spin, and that the necessary 
antisymmetry is provided by the parastatistics. Further, I showed that the
composite operator of three such quarks behaves as a fermion as the nucleon
must. The first sentence of this paper starts with the words 
``Wigner's supermultiplet, ...''\cite{gre}
I redid the magnetic moment calculation using a concrete composite model
for the nucleon with creation operators, all in the $S$-state for the quarks, 
since their parafermi
nature takes care of the antisymmetrization. The calculation is elementary.
The proton with spin up must have quark content $uud$ and spin content
$\uparrow\uparrow\downarrow$. Thus the possible creation operators that
can enter are $u_{\uparrow}$, $u_{\downarrow}$, $d_{\uparrow}$, and 
$d_{\downarrow}$.
For the proton with spin up, the only products of creation operators are
$u_{\uparrow} u_{\uparrow} d_{\downarrow}$ and 
$u_{\uparrow} u_{\downarrow} d_{\uparrow}$.

For the fermi case that Sakita chose, the creation operators anticommute,
\beee
[u,d]_+=0, {\rm etc.}
\eeee
Then the only possibility is
\beee
|p_{\uparrow} \rangle=|u_{\uparrow} u_{\downarrow} d_{\uparrow} \rangle.
\eeee
Then the contributions of the $u$ quarks to the magnetic moment cancel
and the result is
\beee
\mu_p=\mu_d.
\eeee
Since the $d$ has negative charge, the proton magnetic moment is 
antiparallel to its spin; it points the wrong way! The same is true for
the magnetic moment of the neutron.

The bose case (always keeping in mind that the overall antisymmetry is
provided by the parastatistics) is a bit more complicated. Make a neutral
``core'' with $I=S=0$, 
$u_{\uparrow} d_{\downarrow} - u_{\downarrow} d_{\uparrow}$.
Then the proton with spin up is
\beee
|p_{\uparrow}\rangle=\frac{1}{\sqrt{3}}
|u_{\uparrow}(u_{\uparrow} d_{\downarrow} - u_{\downarrow} d_{\uparrow}).
\rangle
\eeee
In terms of the spins carried by the $u$ and $d$ quarks, the magnetic 
moment operator $\mu$ is
\beee
\mu=2 \mu_0 (\frac{2}{3} S_u - \frac {1}{3} S_d),
\eeee
where $\mu_0$ is the Bohr magneton of the quark, and $S_u$ and $S_d$ are the
spin operators of the quarks. The coefficients of the spin operators are
the electric charges of the quarks. Then
\beee
\mu_p=\langle p_{\uparrow}| \mu | p_{\uparrow} \rangle =2 \mu_0 \frac{1}{3} 
\{2[\frac{2}{3}\cdot 1 + (-\frac{1}{3}) \cdot (-\frac{1}{2})]
+(-\frac{1}{3}) \cdot (\frac{1}{2}) \} = \mu_0.
\eeee
Here the $1/3$ factor in front of the curly bracket is the square of the
normalization factor of the proton, the factor $2$ in front of the square
bracket is the bose factor from two $u$ quarks, the fractions of $2/3$ and
$-1/3$ are the charges of the quarks, and the factors of $1$, $-1/2$
and
$1/2$ are the $z$-components of the quark spins. The corresponding result
for the neutron is
\beee
\mu_n= -\frac{2}{3}\mu_0.
\eeee
Thus, as mentioned above,
\beee
\frac{\mu_p}{\mu_n}= -\frac{2}{3}.
\eeee

I pointed out that if the quarks couple minimally, then the mass of the 
quark would be about $1/3$ of the nucleon mass. This is the value that
is now taken for the ``constituent'' mass of the $u$ and $d$ quarks.

Now, thirty eight years after the fact, it is difficult to recover the
mind set of the fall of 1964 when the developments I am describing took
place. At the time several solutions to the statistics paradox were
suggested. These include\\
$\bullet$ Parastatistics, which is equivalent to color as a classification
symmetry\\
$\bullet$ Three integer charged triplets\\
$\bullet$ Quarks are a mathematical fiction\\
$\bullet$ Quarks are indeed fermions with a complicated ground state
wavefunction\\
$\bullet$ Extra quark-antiquark pairs\\
$\bullet$ More exotic possibilities\\

Parastatistics and its equivalence to color as a classification symmetry
is what I am discussing here. Three integer charged triplets were first
suggested by Yoichiro Nambu\cite{nam} and further developed in \cite{nam2}
and \cite{nam3}. Also see \cite{nam4}. I will discuss this case later.
The possibility that quarks are just a mathematical fiction is due to
Gell-Mann in his original paper. This would obviate the statistics paradox
at the cost of making quarks non-physical. Both Nambu and I took an
unequivocal stand that quarks are real and physical. 
Richard H. Dalitz, for some years the rapporteur on hadron
spectroscopy at international conferences, was a leader of those who 
preferred to assume quarks are fermions (without an additional degree of
freedom) and that the statistics paradox would be avoided by having an
antisymmetric space ground state wavefunction. Because there are only 
two (nonrelativistic) space coordinates available when the center-of-mass
coordinate is eliminated, the simplest antisymmetric scalar wavefunction
has degree 6 in the coordinates. This is implausible in view of theorems
that require the ground state wavefunction to be nodeless. In addition,
if the ground state wave function has this complicated form, then there
is no clear candidate for the excited states. As we will see, the
parastatistics theory, in which the ground state wavefunction 
is nodeless,
leads to a simple model for the excited states. Finally, if the ground
state wave function has nodes, then there should be zeroes in the electric
and magnetic form factors of the nucleon. No such zeroes have ever been
found. The possibility of extra quark-antiquark pairs is likely to lead
to ``exploding $SU(3)$ representations,'' which have not been seen. Of
course we now believe that the nucleon and other hadronic states are not
solely described by their constituent quark content; they also have 
terms with extra quark-antiquark pairs as well as gluons; however the
classification of the hadronic states can still be made in terms of the
quantum numbers of their constituent quarks.

The quark statistics is irrelevant for states of mesons. For baryons,
statistics is crucial. I proposed that the baryons should be described
using the ``symmetric'' quark model in which the states are totally
symmetric in terms of the visible space, spin and unitary spin degrees
of freedom\cite{gre}. The antisymmetry of the quarks is taken care
of by the parastatistics of order $3$. 
I developed an atomic model in which the higher
states of baryons have the quarks excited into higher orbital states, 
starting with the $p$ state of opposite parity. 
(Recall that parity was introduced by Wigner, as was time reversal.)
 I gave a table of excited states in my paper of 1964. 
Gabriel Karl and Obryk\cite{kar} later
corrected some of the states.

Since I was at the Institute for Advanced Study, I was eager to hear
the opinion of J. Robert Oppenheimer. I gave him a preprint of my paper
before leaving for the Eastern Theoretical Physics Conference, which we
both attended, that was
to take place at my home university, the University of Maryland in College
Park. When I saw Oppenheimer I asked if he had read my paper. He said

\begin{center}
``Greenberg, your paper is beautiful,''\\
\end{center}

\noindent That made my spirits rise; however he continued

\begin{center}
``but I don't believe a word of it.''\\
\end{center}

\noindent Then my spirits came down again.

I was asked to speak about my paraquark model at Harvard. In the parking
lot after the talk, Julian Schwinger made the prescient remark that the
implicit degree of freedom implicit in the parastatistics model should
play a dynamical role. To my regret I did not follow up on Schwinger's
remark, which pointed toward quantum chromodynamics.

For some years there was resistance to quarks and color.\\
$\bullet$ Unobserved fractionally-charged charged quarks seemed outrageous.\\
$\bullet$ Gell-Mann's ambiguous position cast doubt 
on the reality of quarks.\\
$\bullet$ A new hidden degree of freedom seemed doubly outrageous.

The arguments for quarks and color included baryon and meson spectroscopy,
the magnetic moments of the nucleon, the Zweig rule which predicted 
suppression of decays without connected quark line graphs, the parton
model, the $J/\Psi$ and its friends, and the 
$\pi^0 \rightarrow \gamma \gamma$
decay and the axial anomaly.

Marvin Resnikoff and I gave a detailed fit of the baryons in the 
$({\mathbf 56}, {\mathbf L=0})$ and the 
$({\mathbf 70}, {\mathbf L=1})$ supermultiplets in 1967.\cite{gre2} At the 
International Conference on High Energy Physics (the Rochester Conference)
in Vienna in 1968, 
Haim Harari was the first rapporteur to suggest that the
symmetric quark model is the correct model for baryons. The symmetric 
quark model was developed further by Alvaro De Rujula, Howard Georgi
and Sheldon Glashow\cite{ruj} in the context of the Coulumb potential 
suggested by quantum chromodynamics, by Dalitz and collaborators, and
especially by Nathan Isgur and Karl\cite{isg}. 
Elizabeth Jenkins\cite{jen} gave
a review of baryon spectroscopy from the standpoint of large $N$
quantum chromodynamics. There is surprizing similarity between the old
results of Resnikoff and myself and the large $N$ results.

Nambu was the pioneer in introducing three integer charged triplets and
in having a gauge interaction that couples to the new three-valued degree
of freedom via an octet of what we now call gluons. The proposal to have
three triplets first appears in \cite{nam}. The coupling to the
three-valued degree of freedom is in \cite{nam2} and is developed further
in \cite{nam3}. Nambu and Han gave further analysis in \cite{nam4}.

Quantum chromodynamics has two facets; the hidden three-valued degree of
freedom and the local $SU(3)$ gauge theory with a vector octet of 
self-interacting gluons that mediate the forces between the particles
that carry the degree of freedom. These two facets are analogous to the
electric charge as a degree of freedom and to the $U(1)$ gauge theory 
in which photons mediate the interaction between charged particles 
in electromagnetism. The three-valued degree of freedom is
implicit in the paraquark model, which in equivalent to color as a
classification symmetry. The second facet of quantum chromodynamics is
the local $SU(3)$ color gauge interaction that provides the octet of
gluons that mediates the force between color-carrying particles. These
two facets taken together constitute quantum chromodynamics.

The word ``color'' that is colloquially attached to this new degree of 
freedom was introduced first by Pais\cite{pai} in a lecture at the Erice Summer
School in 1965. Donald B. Lichtenberg\cite{lic} used the word independently in his
book in 1970. Color became used generally after the papers of 
Gell-Mann\cite{gel2} and of William A. Bardeen, Harald Fritzsch and 
Gell-Mann\cite{bar}.

Some of the effects connected with color come from color as a classification
symmetry. These include the classification of states of baryons and of 
mesons according to their constituent quark structure, the 
$\pi^0 \rightarrow \gamma \gamma$ decay rate that follows from the axial
anomaly, and the ratio of the electron-positron annihilation cross section
to hadrons to the corresponding cross section to muon pairs. In each case, the number of
quarks circulating in the quark loop gives the necessary factor $N$ to agree
with experiment. This $N$ is equivalently 
either the number of Green components in
the paraquark theory or the number of colors in the $SU(3)$ color theory. 
Other effects require the gauged theory of color, including asymptotic
freedom which allows the reconciliation of the constituent quark model for
static properties of hadrons with the parton model for high-energy 
collision processes, as well as the observed running of the strong coupling
constant, confinement, and the observed two-gluon and 
three-gluon jets.

The acceptance of the quark model occurred over a period of years. In
the period 1964-1966, the main evidence was the baryon spectra, the 
magnetic moment ratio $\mu_p/\mu_n$, the Zweig rule, relations among
cross sections, such as $\sigma_{\pi N}/\sigma_{NN}=2/3$ that follow from
simple quark counting arguments. Later the $\pi^0$ decay and the 
electron-positron annihilation cross section ratio provided support. In
1969 the SLAC deep inelastic electron scattering experiments as interpreted
using Bjorken scaling, Feynman's parton model, and the Bjorken-Paschos
quark-parton model greatly strengthened the case. However only in 1974,
with the discovery of the $J/\Psi$ and its friends as bound states of
charm and anticharm quarks, did the quark model
with color become accepted generally.

Quantum chromodynamics can be summarized in terms of its gauge Lagrangian
for $SU(3)$ color with three generations of color-triplet 
quark matter fields. The reason for three generations is beyond the scope
of quantum chromodynamics and is not yet known. The main features of
quantum chromodynamics as a gauge theory are the running of the coupling constant and the
associated asymptotic freedom at high energy and infrared slavery which
leads to permanent confinement of color-carrying particles (at zero
temperature).

QCD has passed many tests, including the running of $\alpha_{strong}$,
the jets in hadronic collisions, the (modified) scaling of scattering
processes, and the parton model structure and fragmentation functions.
QCD agrees well with data on heavy quarkonium decays, electron-positron
annihilation to hadrons and to leptons, and with the measurement of the
$Z$ width, which gives a restriction on the number of neutrinos with
mass below half the mass of the $Z$.

I conclude with cursory remarks about the theoretical analysis of QCD.
Broadly speaking, there are three approaches, (1) models that don't try
to start from first principles, i.e. from the QCD Lagrangian. There are
many models, none universally accepted, (2) continuum methods, mainly
using Bethe-Salpeter methods, which have
had, at best, limited success, and (3) lattice QCD, pioneered by Kenneth
Wilson, which at present seems to be the most productive approach, and, at
least, has stimulated development of supercomputers.

This whole story of the progress from Wigner's supermultiplet theory of
nuclei to quantum chromodynamics illustrates Wigner's profound influence
on the physics of his time and beyond: the centrality of symmetry
principles, the use of deep general mathematical results, and the direct
contact with physical phenomena.

\end{document}